\documentclass [12pt]{article}
\usepackage {amssymb}
\usepackage {amsmath}
\usepackage[dvips]{epsfig}

\begin{document}
\begin{center}

\Large\textbf{Analytical solutions for the interstitial diffusion
of impurity atoms}
\\[2ex]
\normalsize
\end{center}

\begin{center}
\textbf{O. I. Velichko, N. A. Sobolevskaya}
\end{center}

\begin{center}
\textit{Belarusian State University of Informatics and
Radioelectronics}

Department of Physics, Belarusian State University of Informatics
and Radioelectronics, 6, P.~Brovki Street, Minsk, 220013 Belarus

{\it E-mail addresses: oleg\_velichko@lycos.com} (Oleg Velichko)

{\it sobolevskaya@lycos.com} (Natalia Sobolevskaya)
\end{center}

Abstract

The analytical solutions of the equations describing impurity
diffusion due to migration of nonequilibrium impurity
interstitials were obtained for the impurity redistribution during
ion implantation at elevated temperatures and for diffusion from a
doped epitaxial layer. The reflecting boundary condition at the
surface of a semiconductor and the conditions of  constant
concentrations at the surface and in the bulk of it were used in
the first and second cases, respectively. On the basis of these
solutions  hydrogen diffusion in silicon during high-fluence
low-energy deuterium implantation and beryllium diffusion from a
doped epi-layer during rapid thermal annealing of InP/InGaAs
heterostructures were investigated. The calculated impurity
concentration profiles agree well with experimental data. The
fitting to the experimental profiles allowed us to derive the
values of the parameters that describe interstitial impurity
diffusion.

\bigskip
{\it PACS:} 61.72.Tt;66.30.Dn; 66.30.Jt; 02.60.Cb

{\it Keywords:} implantation; diffusion; equation solution;
interstitial; hydrogen; beryllium

\bigskip

\section{Introduction}

The recent years numerical methods have been widely used for
simulation of solid state diffusion of ion-implanted dopants (see,
for example, \cite{Simulator_00,Temkin_05}). As a rule, to
simulate the impurity diffusion a system of equations describing a
coupled diffusion of different mobile species and their
quasichemical reactions during annealing is solved. Due to a great
number of differential equations and the complexity of the system
as a whole, the problem of the correctness of a numerical solution
is very important. One of the best ways to verify the correctness
of an approximate numerical solution is a comparison with the
exact analytical solution of the boundary value problem under
consideration. Such analytical solutions can be derived for the
special simplest cases of dopant or point defect diffusion
processes. For example, in Ref. \cite{Minear_72} an analytical
solution for the point defect diffusion based on the method of
Green's functions was obtained. It was supposed in
\cite{Minear_72} that nonequilibrium point defects were
continuously generated during ion implantation of impurity atoms
and diffused to the surface and into the bulk of a semiconductor.
The surface was considered to be a perfect sink for point defects.
In Ref. \cite{Velichko_88} a process of impurity diffusion during
ion implantation at elevated temperatures was investigated
analytically. It was supposed that the implantation temperature
was too low to provide a traditional diffusion by the ``dopant
atom -- point defect" pairs, but was enough for the diffusion of
nonequilibrium interstitial impurity atoms to occur. Unlike
\cite{Minear_72}, in
 Ref. \cite{Velichko_88} a system of equations, namely, the
conservation law for substitutionally dissolved impurity atoms and
equation of diffusion--recombination of nonequilibrium
interstitial impurity atoms have been solved analytically by the
method of Green's functions. Reflecting boundary condition at the
surface of a semiconductor has been chosen to describe the
interaction of interstitial impurity atoms with the interface. Due
to this condition, a diffusion problem has become symmetric with
respect to the point $x=0$. For simplicity, the condition of zero
impurity concentration for $x \rightarrow \pm \infty$ has been
used. It is interesting to note that analytical solutions for gold
diffusion in silicon due to Frank-Turnbull and due to kick-out
mechanisms were obtained in Ref. \cite{Goesele_80} and Refs.
\cite{Goesele_80,Seeger_80}, respectively. It was supposed that
there was a local equilibrium between substitutionally dissolved
impurity atoms, vacancies (or self-interstitials for the kick-out
mechanism) and interstitial impurity atoms. The case of
nonequilibrium interstitial impurity atoms was not considered in
these papers. The very interesting case of coupled diffusion of
vacancies and self-interstitials was investigated in
\cite{Hashimoto_90,Okino_95}. The equations of the diffusion of
vacancies or of self-interstitials  are similar to the equation of
diffusion of impurity interstitials. However, the solutions
obtained in \cite{Hashimoto_90,Okino_95} are difficult to use for
describing the impurity diffusion governed by nonequilibrium
interstitial impurity atoms, because a condition of local
equilibrium was used in all these papers. Besides, a generation
rate was assumed to be equal to zero in \cite{Okino_95} or equal
to constant value in \cite{Hashimoto_90}. Thus, it is reasonable
to derive an analytical solution for the case of diffusion due to
migration of nonequilibrium impurity interstitials.

The main goal of this paper is to continue the investigation
\cite{Velichko_88} to obtain  other analytical solutions and
compare these solutions with the experimental data.

\section{Original equations}

It is supposed that the processing temperature is too low to
provide a diffusion of substitutionally dissolved impurity atoms,
but is enough for the diffusion of impurity interstitials. The
generation of nonequilibrium interstitial impurity atoms can occur
due to ion implantation, including the case of ion implantation at
elevated temperatures, or due to the replacement of the impurity
atom by self-interstitial from the substitutional position to the
interstitial one (Watkins effect \cite{Watkins_69}), or due to
dissolution of the clusters that incorporate impurity atoms etc.
It is also supposed that the impurity concentration in the doped
regions formed due to migration of nonequilibrium impurity
interstitials is smaller than $n_{2}$  or approximately equal to
$n_{i}$ or that impurity atoms in interstitial position are
neutral. Here $n_{i}$ is the intrinsic carrier concentration at
the processing temperature. Then, the system of equations
describing the evolution of impurity concentration profiles
includes \cite{Velichko_88}:

(i) a conservation law for substitutionally dissolved impurity
atoms:

\begin{equation}\label{Conservation law}
\displaystyle \frac{\partial \, C(x,t)}{ \partial \, t} =
\displaystyle \frac{C^{AI}(x,t)}{\tau^{AI}} + G^{AS}(x,t)\, ,
\end{equation}

(ii) an equation of diffusion for nonequilibrium interstitial
impurity atoms:

\begin{equation}\label{Nonequilibrium impurity interstitials}
d^{AI}\displaystyle \frac{\partial^{2} \, C^{AI}}{ \partial \,
x^{2}} - \displaystyle \frac{C^{AI}}{\tau^{AI}} + G^{AI}(x,t) = 0
\, ,
\end{equation}

or

\begin{equation}\label{Normalized equation}
- \left[ \displaystyle \frac{\partial^{2} \, C^{AI}}{ \partial \,
x^{2}} - \displaystyle \frac{C^{AI}}{l_{AI}^{2}}\right] =
\frac{\tilde{g}^{AI}(x,t)}{l_{AI}^{2}} \, ,
\end{equation}

where

\begin{equation}\label{Average migration lenght}
l_{AI}=\sqrt{d^{AI} \tau^{AI}} \, , \qquad \tilde{g}^{AI}(x,t)=
G^{AI}(x,t) \, \tau^{AI} \, .
\end{equation}

Here $C$ and $C^{AI}$ are the concentrations of substitutionally
dissolved impurity atoms and nonequilibrium impurity
interstitials, respectively; $G^{AS}$ is the rate of introducing
of impurity atoms, which immediately occupy the substitutional
positions; $d^{AI}$ and $\tau^{AI}$ are the diffusivity and
average lifetime of nonequilibrium interstitial impurity atoms,
respectively; $G^{AI}$ is the generation rate of interstitial
impurity atoms. We use a steady-state diffusion equation for
impurity interstitials, because of the large average migration
length of nonequilibrium interstitial impurity atoms ($l_{AI} \gg
l_{fall}$, where $l_{fall}$ is the characteristic length of the
decrease in the impurity concentration) and due to the small
average lifetime of nonequilibrium impurity interstitials ($
\tau_{AI} \ll t_{P}$, where $t_{P}$ is the duration of thermal
treatment).

The system (\ref{Conservation law}), (\ref{Nonequilibrium impurity
interstitials}) or (\ref{Conservation law}), (\ref{Normalized
equation}) describes impurity diffusion due to migration of
nonequilibrium interstitial impurity atoms. To solve this system
of equations, appropriate boundary conditions are need. Let us
consider, in contrast to \cite{Velichko_88}, the finite-length
one-dimensional (1D) domain $[0,x_{B}]$, i.e., the domain used in
1D numerical modeling, and add the following boundary conditions
to Eq. (\ref{Normalized equation}):

\begin{equation}\label{Boundary left}
\mathrm{w}^{S}_{1}d^{AI}\left. \frac{\partial  \, C^{AI}}{\partial
x} \right |_{\displaystyle x=0} +\mathrm{w}^{S}_{2} \left. C^{AI}
\right |_{\displaystyle x=0} = \mathrm{w}^{S}_{3} \, ,
\end{equation}

\begin{equation}\label{Boundary right}
\mathrm{w}^{B}_{1}d^{AI}\left. \frac{\partial  \, C^{AI}}{\partial
x} \right |_{\displaystyle x=x_{B}} +\mathrm{w}^{B}_{2} \left.
C^{AI} \right |_{\displaystyle x=x_{B}} = \mathrm{w}^{B}_{3} \, ,
\end{equation}

as well as the initial conditions:

\begin{equation}\label{Initial_conditions}
C(x,0)=C_{0}(x) \, , \qquad  C^{AI}(x,0)=C^{AI}_{eq}=const \, .
\end{equation}

Here, $\mathrm{w}^{S}_{1}$, $\mathrm{w}^{S}_{2}$,
$\mathrm{w}^{S}_{3}$ and $\mathrm{w}^{B}_{1}$,
$\mathrm{w}^{B}_{2}$, $\mathrm{w}^{B}_{3}$ are  constant
coefficients; $C^{AI}_{eq}$ is the equilibrium value of
concentration of interstitial impurity atoms (it is supposed that
$C^{AI}_{eq}$ is equal to zero for many cases under
consideration).

To derive an analytical solution of this boundary value problem,
the method of Green's function \cite{Butkovskiy_83} can be used.

\section{Analytical method and solutions}

The suggestion about the immobile substitutionally dissolved
impurity atoms allows one to solve Eq. (\ref{Conservation law})
independently of Eq. (\ref{Nonequilibrium impurity interstitials})
or Eq. (\ref{Normalized equation}):

\begin{equation}\label{Main integral}
C(x,t) = \frac{1}{\tau^{AI}}\int \limits_{0}^{\displaystyle
t}C^{AI}(x,t)dt +\int \limits_{0}^{\displaystyle t}G^{AS}(x,t)dt +
C_{0}(x) \, .
\end{equation}

We will supplement expression (\ref{Main integral}) with a
steady-state solution of Eq. (\ref{Normalized equation}) obtained
by the method of Green's functions \cite{Butkovskiy_83}:

\begin{equation}\label{Steady-state_solution}
C^{AI}(x,t) = \int \limits_{0}^{\displaystyle
x_{B}}G(x,\xi)w(\xi,t)d\xi \, ,
\end{equation}

where

\begin{equation}\label{Standard_function}
w(\xi,t) = \frac{\tilde{g}^{AI}(\xi,t)}{l_{AI}^{2}} + w_{S}(\xi) +
w_{B}(\xi) \, .
\end{equation}

Here $G(x,\xi)$ is the Green's function for Eq. (\ref{Normalized
equation}). Using the standardizing function $w(x,t)$
\cite{Butkovskiy_83} allows one to reduce the previous boundary
value problem to the boundary value problem with zero boundary
conditions:

\begin{equation}\label{Boundary left_zero}
\mathrm{w}^{S}_{1}d^{AI}\left. \frac{\partial  \, C^{AI}}{\partial
x} \right |_{\displaystyle x=0} +\mathrm{w}^{S}_{2} \left. C^{AI}
\right |_{\displaystyle x=0} = 0 \, ,
\end{equation}

\begin{equation}\label{Boundary right_zero}
\mathrm{w}^{B}_{1}d^{AI}\left. \frac{\partial  \, C^{AI}}{\partial
x} \right |_{\displaystyle x=x_{B}} +\mathrm{w}^{B}_{2} \left.
C^{AI} \right |_{\displaystyle x=x_{B}} = 0 \, .
\end{equation}

The Green's function for Eq. (\ref{Normalized equation}) with
boundary conditions (\ref{Boundary left_zero}) and (\ref{Boundary
right_zero}) has the following form \cite{Butkovskiy_83}:

\begin{equation}\label{Green's_function}
   G(x,\xi)= \frac{1}{K}\left \{ \begin{array}{cc}
Q_{1}(x)Q_{2}(\xi) & \mbox{\ for } 0\leq x \leq \xi \leq x_{B} \, ,\\
Q_{1}(\xi)Q_{2}(x) & \mbox{\ for } 0\leq \xi \leq x \leq x_{B} \,
,
\end{array} \right.
\end{equation}

\noindent where

\begin{equation}\label{K}
K= Q_{1}^{'}(x)Q_{2}(x)-Q_{1}(x)Q_{2}^{'}(x)=const \, .
\end{equation}

Here $Q_{1}$ and $Q_{2}$ are the linearly independent solutions of
the homogeneous equation

\begin{equation}\label{Homogeneous equation}
\displaystyle \frac{\mathrm{d}^{2} \, Q}{\mathrm{d} \, x^{2}} -
\displaystyle \frac{Q}{l_{AI}^{2}} = 0 \,
\end{equation}

\noindent with the following initial conditions on the left
boundary:

\begin{equation}\label{Boundary Q1}
Q_{1}(0)= \mathrm{w}^{S}_{1}d^{AI} \, , \qquad Q^{'}_{1}(0)=
-\mathrm{w}^{S}_{2} \, ,
\end{equation}

\noindent and on the right one:

\begin{equation}\label{Boundary Q2}
Q_{2}(x_{B})= \mathrm{w}^{B}_{1}d^{AI} \, , \qquad
Q^{'}_{2}(x_{B})= -\mathrm{w}^{B}_{2} \, .
\end{equation}

Following \cite{Butkovskiy_83}, one can write the functions
$w_{S}(x)$ and $w_{B}(x)$ as

\begin{equation}\label{wL}
   w_{S}(x)= \left \{ \begin{array}{cc}
\displaystyle
-\frac{1}{\mathrm{w}^{S}_{1}d^{AI}}\delta(-x)\mathrm{w}^{S}_{3} &
\mbox{\ if }  \, \mathrm{w}^{S}_{1} \neq 0 \, ,\\
\\
\displaystyle
\frac{1}{\mathrm{w}^{S}_{2}}\delta^{'}(-x)\mathrm{w}^{S}_{3} &
\mbox{\ if }  \, \mathrm{w}^{S}_{2} \neq 0 \, ,
\end{array} \right.
\end{equation}

\begin{equation}\label{wR}
   w_{B}(x)= \left \{ \begin{array}{cc}
\displaystyle
\frac{1}{\mathrm{w}^{B}_{1}d^{AI}}\delta(x_{B}-x)\mathrm{w}^{B}_{3}
&
\mbox{\ if }  \, \mathrm{w}^{B}_{1} \neq 0 \, ,\\
\\
\displaystyle
-\frac{1}{\mathrm{w}^{B}_{2}}\delta^{'}(x_{B}-x)\mathrm{w}^{B}_{3}
& \mbox{\ if }  \, \mathrm{w}^{B}_{2} \neq 0 \, .
\end{array} \right.
\end{equation}

Thus, for the reflecting boundary condition

\begin{equation}\label{Boundary_Case1 left}
\mathrm{w}^{S}_{1}=1, \qquad \mathrm{w}^{S}_{2}=0, \qquad
\mathrm{w}^{S}_{3}=0 \,
\end{equation}

\noindent at the surface of a semiconductor ($x=0$) and for
Dirichlet boundary condition

\begin{equation}\label{Boundary_Case1 right}
\mathrm{w}^{B}_{1}=0, \qquad \mathrm{w}^{B}_{2}=1, \qquad
\mathrm{w}^{B}_{3}=C^{AI}_{B}
\end{equation}

\noindent in the bulk ($x=x_{B}$) the solutions $Q_{1}$ and
$Q_{2}$ have the form

\begin{equation}\label{Homogeneous solutions}
Q_{1}(x)= d^{AI} \cosh \left (\frac{x}{l_{AI}} \right ) \, ,
\qquad Q_{2}(x)= l_{AI} \sinh \left (\frac{x_{B}-x}{l_{AI}} \right
) \, ,
\end{equation}

\begin{equation}\label{K1}
K= d^{AI}\cosh \left( \frac{x_{B}}{l_{AI}} \right)=const \, .
\end{equation}

\begin{equation}\label{Green's_function1}
   G(x,\xi)= \frac{ l_{AI}}{\displaystyle
   \cosh \left( \frac{x_{B}}{l_{AI}} \right)}
   \left \{ \begin{array}{cc}
 \cosh \left( \displaystyle \frac{x}{l_{AI}} \right )
 \sinh \left( \displaystyle \frac{x_{B}-\xi}{l_{AI}} \right )
 \mbox{\ for } 0\leq x \leq \xi \leq x_{B} \, ,\\
\\
 \cosh \left( \displaystyle \frac{\xi}{l_{AI}} \right )
 \sinh \left( \displaystyle \frac{x_{B}-x}{l_{AI}} \right )
\mbox{\ for } 0\leq \xi \leq x \leq x_{B} \, ,
\end{array} \right.
\end{equation}

\begin{equation}\label{wL_wR}
w_{S}(x)= 0, \qquad w_{B}(x)= \delta^{'}(x_{B}-x)C^{AI}_{B} \, .
\end{equation}

Let us consider the process of ion implantation in a semiconductor
at an elevated temperature (the so-called ``hot" implantation). It
is established experimentally that the main part of the implanted
impurity atoms occupies  substitutional positions near the places
where they were stopped. Let us suppose that the remaining
nonequilibrium atoms occupy interstitial sites and can diffuse
during implantation before they transfer to the substitutional
position. To describe the spatial distributions of both impurity
atoms directly occupying substitutional position and impurity
interstitials generated during implantation, Gaussian
distributions can be used:

\begin{equation}\label{Generation}
G^{AS}(x,t)= g_{m}(1-p^{AI})\exp \left[
-\frac{(x-R_{p})^{2}}{2\triangle R_{p}^{ \,2}}\right] \, ,
\end{equation}

\begin{equation}\label{Interstitial_Generation}
G^{AI}(x,t)= g_{m}p^{AI}\exp \left[
-\frac{(x-R_{p})^{2}}{2\triangle R_{p}^{ \,2}}\right] \, ,
\end{equation}

\noindent where $g_{m}$ is the number of impurity atoms being
introduced per unit area per second by ion implantation; $p^{AI}$
is a part of the implanted impurity atoms occupying interstitial
sites; $R_{p}$ and $\triangle R_{p}$ are the average projective
range of implanted ions and straggling of projective range,
respectively.

Taking into consideration  expressions (\ref{wL_wR}) and
(\ref{Interstitial_Generation}) yields

\begin{equation}\label{Standard_function1}
w(\xi,t) = \frac{g_{m}p^{AI}\tau^{AI}}{l_{AI}^{2}}\exp \left[
-\frac{(\xi-R_{p})^{2}}{2\triangle R_{p}^{ \,2}}\right] -
\delta^{'}(x_{B}-\xi)C^{AI}_{B} \, .
\end{equation}

Substituting Green's function (\ref{Green's_function1}) and
standardizing function (\ref{Standard_function1}) into expression
(\ref{Steady-state_solution}) allow one to obtain a spatial
distribution of diffusing interstitial impurity atoms:

\begin{equation}\label{Steady-state_solution1}
\begin {array} {c}
C^{AI}(x,t) = \int \limits_{\displaystyle 0}^{\displaystyle
x_{B}}G(x,\xi)w(\xi,t)d\xi  = \displaystyle
\frac{g_{m}p^{AI}\tau^{AI}}{l_{AI}}
\cosh^{-1}\left(\frac{x_{B}}{l_{AI}} \right) \\
 \\
 \times \left\{ \sinh\left( \displaystyle
\frac{x_{B}-x}{l_{AI}}\right) \int \limits_{\displaystyle
0}^{\displaystyle x} \cosh\left( \displaystyle
\frac{\xi}{l_{AI}}\right) \exp \left[\displaystyle
-\frac{(R_{p}-\xi)^{2}}{2\triangle R_{p}^{
\,2}}\right] d\xi \right. \\
 \\
 \left. +\cosh\left( \displaystyle
\frac{x}{l_{AI}}\right) \int \limits_{\displaystyle
x}^{\displaystyle x_{B}} \sinh\left( \displaystyle
\frac{x_{B}-\xi}{l_{AI}}\right) \exp \left[\displaystyle
-\frac{(R_{p}-\xi)^{2}}{2\triangle R_{p}^{ \,2}}\right] d\xi
\right\} \\
 \\
 + C^{AI}_{B}\cosh^{-1}\left( \displaystyle \frac{x_{B}}{l_{AI}}\right)
 \cosh\left( \displaystyle \frac{x}{l_{AI}}\right)
\int \limits_{\displaystyle x}^{\displaystyle x_{B}} \sinh\left(
\displaystyle \frac{\xi-x_{B}}{l_{AI}}\right)
\delta^{'}(x_{B}-\xi) d\xi \, .
\end {array}
\end{equation}

Calculating the integrals on the right-hand side of expression
(\ref{Steady-state_solution1}) one can obtain an explicit
expression for the distribution of interstitial impurity atoms:

\begin{equation}\label{Steady-state_solution1D}
\begin {array} {c}
C^{AI}(x,t)= C_{m} \displaystyle \frac{\exp \, u_{1}}{\cosh
 \, u^{B}_{2}} \left\{ \cosh
u_{2} [\exp(-u_{6})
(\mathrm{erf} u^{B}_{4}-\mathrm{erf} u_{4})\right . \\
\\
+\exp(u_{6})(\mathrm{erf} u^{B}_{5}-\mathrm{erf} u_{5})]
 +\exp(-u_{9})\sinh u_{3} \\
\\

\left . \times [\mathrm{erf} u_{7}+\exp(2u_{9})(\mathrm{erf}
u_{5}- \mathrm{erf} u_{8})-\mathrm{erf} u_{4}] \right \}
+C^{AI}_{B} \displaystyle \frac{\cosh u_{2}}{\cosh u^{B}_{2}} \, ,
\end {array}
\end{equation}

\noindent where

\begin{equation}\label{Cm}
C_{m}=\displaystyle \frac{\sqrt{\pi}g_{m}p^{AI}\tau^{AI}\Delta
R_{p}}{2\sqrt{2}\, l_{AI}}
 \, ,
\end{equation}

\begin{equation}\label{u1}
u_{1}= \frac{\Delta R^{\,2}_{p}}{2l^{\,2}_{AI}} \, ,
\end{equation}

\begin{equation}\label{u2}
u_{2}= \frac{x}{l_{AI}}  \, , \qquad  u^{B}_{2}=
\frac{x_{B}}{l_{AI}} \, ,
\end{equation}

\begin{equation}\label{u3}
 u_{3}= \frac{x-x_{B}}{l_{AI}} \, ,
\end{equation}

\begin{equation}\label{u4}
u_{4}= \frac{\Delta
R^{\,2}_{p}-l_{AI}R_{p}+l_{AI}x}{\sqrt{2}\Delta R_{p}l_{AI}} \, ,
\qquad u^{B}_{4}= \frac{\Delta
R^{\,2}_{p}-l_{AI}R_{p}+l_{AI}x_{B}}{\sqrt{2}\Delta R_{p}l_{AI}}
\, ,
\end{equation}

\begin{equation}\label{u5}
u_{5}= \frac{\Delta
R^{\,2}_{p}+l_{AI}R_{p}-l_{AI}x}{\sqrt{2}\Delta R_{p}l_{AI}} \, ,
\qquad u^{B}_{5}= \frac{\Delta
R^{\,2}_{p}+l_{AI}R_{p}-l_{AI}x_{B}}{\sqrt{2}\Delta R_{p}l_{AI}}
\, ,
\end{equation}

\begin{equation}\label{u6}
 u_{6}= \frac{R_{p}-x_{B}}{l_{AI}} \, ,
\end{equation}

\begin{equation}\label{u7u8}
u_{7}= \frac{\Delta R^{\,2}_{p}-l_{AI}R_{p}}{\sqrt{2}\Delta
R_{p}l_{AI}} \, , \qquad
u_{8}= \frac{\Delta
R^{\,2}_{p}+l_{AI}R_{p}}{\sqrt{2}\Delta R_{p}l_{AI}} \, ,
\end{equation}

\begin{equation}\label{u9}
u_{9}= \frac{R_{p}}{l_{AI}} \, .
\end{equation}

In a similar way, one can obtain a solution for the case of
impurity interstitial recombination at the surface of a
semiconductor. Let us consider, for example, a buried layer
uniformly doped by impurity atoms. If the impurity concentration
is high, generation of nonequilibrium interstitial impurity atoms
is possible within this layer during thermal treatment. The
boundary conditions assuming the constant impurity interstitial
concentrations at the surface and in the bulk of a semiconductor
(Dirichlet boundary conditions) can be enforced to describe the
interstitial migration

\begin{equation}\label{Boundary left_right}
\left. C^{AI} \right |_{\displaystyle x=0} = C^{AI}_{S} \, ,
\qquad \left. C^{AI} \right |_{\displaystyle x=x_{B}} =C^{AI}_{B}
\, ,
\end{equation}

\noindent where $C^{AI}_{S}$ is the concentration of interstitial
impurity atoms at the surface (it is quite likely that
$C^{AI}_{S}=0$).

Let us suppose that the generation of impurity interstitials is
described by the following function:

\begin{equation}\label{Generation2}
   G^{AI}(x)=
   \left \{ \begin{array}{cc}
0 \qquad \mbox{\ for } 0\leq x  < x_{L} \, ,\\
\\
g_{m} \qquad \, \mbox{\ for } x_{L} \leq x \leq x_{R} \, ,\\
\\
 \, \, \, 0 \qquad \mbox{\ for } x_{R} < x \leq x_{B} \, ,
\end{array} \right.
\end{equation}

\noindent where $g_{m}=const$ and $x_{L}$ and $x_{R}$ are the left
and right boundaries of the doped layer, respectively.

Then, the Green's function and standardizing function are
respectively

\begin{equation}\label{Green's_function2}
   G(x,\xi)= \frac{l_{AI}}{\displaystyle
   \sinh \left( \frac{x_{B}}{l_{AI}} \right)}
   \left \{ \begin{array}{cc}
 \sinh \left( \displaystyle \frac{x}{l_{AI}} \right )
  \sinh \left( \displaystyle \frac{x_{B}-\xi}{l_{AI}} \right )
 \mbox{\ for } 0\leq x \leq \xi \leq x_{B} \, ,\\
\\
 \sinh \left( \displaystyle \frac{\xi}{l_{AI}} \right )
 \sinh \left( \displaystyle \frac{x_{B}-x}{l_{AI}} \right )
\mbox{\ for } 0\leq \xi \leq x \leq x_{B} \, ,
\end{array} \right.
\end{equation}

\noindent and

\begin{equation}\label{Standard_function2}
w(\xi,t) = \frac{G_{AI}(\xi) \, \tau^{AI}}{l_{AI}^{2}}
-\delta^{'}(-\xi)C^{AI}_{S} -\delta^{'}(x_{B}-\xi)C^{AI}_{B} \, .
\end{equation}

Substituting (\ref{Generation2}), (\ref{Green's_function2}), and
(\ref{Standard_function2}) into expression
(\ref{Steady-state_solution}), we obtain a spatial distribution of
the diffusing interstitial impurity atoms:

\begin{equation}\label{Steady-state_solution2}
C^{AI}(x,t) = C^{AI}(x) = C^{AI}_{p}(x)+C^{AI}_{h}(x) \, ,
\end{equation}

\noindent where

\begin{equation}\label{Partial_solution}
\begin {array} {c}
C^{AI}_{p}(x)= \displaystyle \frac{\tau^{AI}}{l_{AI}} \sinh^{-1}
\left( \displaystyle \frac{x_{B}}{l_{AI}}
 \right )  \left[\sinh
\left( \displaystyle \frac{x_{B}-x}{l_{AI}}
 \right ) \int \limits_{\displaystyle 0}^{\displaystyle x}
\sinh \left(\displaystyle \frac{\xi}{l_{AI}} \right )G_{AI}(\xi)
d\xi \right . \\
 \\
 \qquad \qquad \left . + \sinh \left( \displaystyle
 \frac{x}{l_{AI}} \right )  \displaystyle
 \int \limits_{\displaystyle x}^{\displaystyle x_{B}} \sinh \left(
 \displaystyle \frac{x_{B}-\xi}{l_{AI}} \right ) G_{AI}(\xi)
 d\xi \right ] \, ,
\end {array}
\end{equation}

\begin{equation}\label{Homogeneous_solution}
\begin {array} {c}
C^{AI}_{h}(x) =  - C^{AI}_{S}\int \limits_{\displaystyle
0}^{\displaystyle x_{B}} G(x,\xi) \delta^{'}(-\xi) d\xi -
C^{AI}_{B}\int \limits_{\displaystyle 0}^{\displaystyle x_{B}}
G(x,\xi) \delta^{'}(x_{B}-\xi) d\xi
\\
\qquad \qquad = \sinh^{-1}\left(\displaystyle
\frac{x_{B}}{l_{AI}}\right) \left[ C^{AI}_{S} \sinh
\left(\displaystyle \frac{x_{B}-x}{l_{AI}} \right)+C^{AI}_{B}
\sinh \left(\displaystyle \frac{x}{l_{AI}} \right) \right] \, .
\end {array}
\end{equation}

To calculate the integrals in  expression
(\ref{Partial_solution}), let us consider the  following three
cases:

\noindent i$\left. \right)$ If $x<x_{L}$, then

\begin{equation}\label{Integral_1}
\begin {array} {c}
C^{AI}_{p}(x) =\displaystyle \frac{\tau^{AI}g_{m}}{l_{AI}} \sinh
\left( \displaystyle
 \frac{x}{l_{AI}} \right )  \displaystyle
 \int \limits_{\displaystyle x_{L}}^{\displaystyle x_{R}} \sinh \left(
 \displaystyle \frac{x_{B}-\xi}{l_{AI}} \right ) d\xi  =\tau^{AI} g_{m}  \\
 \\
\times \sinh^{-1} \left( \displaystyle \frac{x_{B}}{l_{AI}}
\right) \sinh \left( \displaystyle \frac{x}{l_{AI}}\right) \left[
\cosh \left( \displaystyle \frac{x_{B}-x_{L}}{l_{AI}} \right) -
\cosh \left( \displaystyle \frac{x_{B}-x_{R}}{l_{AI}} \right)
\right] \, ,
\end {array}
\end{equation}

\noindent ii$\left. \right)$ If $x_{L}\leq x \leq x_{R}$, then

\begin{equation}\label{Integral_2}
\begin {array} {c}
C^{AI}_{p}(x) = \displaystyle \frac{g_{m} \,\tau^{AI}}{l_{AI}}
\sinh^{-1} \left( \displaystyle \frac{x_{B}}{l_{AI}}
 \right ) \left[  \sinh \left( \displaystyle
 \frac{x_{B}-x}{l_{AI}} \right ) \displaystyle
 \int \limits_{\displaystyle x_{L}}^{\displaystyle x} \sinh \left(
 \displaystyle \frac{\xi}{l_{AI}} \right ) d\xi +  \right. \\
\\
 \left. + \sinh
\left( \displaystyle \frac{x}{l_{AI}}
 \right ) \displaystyle \int
 \limits_{\displaystyle x}^{\displaystyle x_{R}}
\sinh \left(\displaystyle \frac{x_{B}-\xi}{l_{AI}} \right ) d\xi \right]\\
\\
  = g_{m} \,\tau^{AI} \sinh^{-1} \left( \displaystyle
\frac{x_{B}}{l_{AI}} \right) \left\{\sinh \left( \displaystyle
\frac{x_{B}-x}{l_{AI}}\right) \left[ \cosh \left( \displaystyle
\frac{x}{l_{AI}} \right) - \cosh \left( \displaystyle
\frac{x_{L}}{l_{AI}}
\right)  \right] \right . \\
\\
\left .  + \sinh \left( \displaystyle \frac{x}{l_{AI}}\right)
\left[\cosh \left( \displaystyle \frac{x_{B}-x}{l_{AI}} \right) -
\cosh \left( \displaystyle \frac{x_{B}-x_{R}}{l_{AI}} \right)
  \right] \right\} \, ,
\end {array}
\end{equation}

\noindent iii$\left. \right)$ If $x > x_{R}$, then

\begin{equation}\label{Integral_3}
\begin {array} {c}
C^{AI}_{p}(x)= \displaystyle \frac{\tau^{AI}g_{m}}{l_{AI}} \sinh
\left( \displaystyle \frac{x_{B}-x}{l_{AI}}
 \right ) \int \limits_{\displaystyle x_{L}}^{\displaystyle x_{R}}
\sinh \left(\displaystyle \frac{\xi}{l_{AI}} \right )
d\xi  \\
 \\
=\tau^{AI} g_{m} \sinh^{-1} \left( \displaystyle
\frac{x_{B}}{l_{AI}} \right) \sinh \left( \displaystyle
\frac{x_{B}-x}{l_{AI}}\right) \left[ \cosh \left( \displaystyle
\frac{x_{R}}{l_{AI}} \right) - \cosh \left( \displaystyle
\frac{x_{L}}{l_{AI}} \right)  \right]  \, .
\end {array}
\end{equation}

Expressions (\ref{Integral_1}), (\ref{Integral_2}), and
(\ref{Integral_3}) are the partial steady-state solution of the
boundary value problem under consideration for the case of step
distribution of interstitial generation rate (\ref{Generation2})
and zero concentrations of nonequilibrium impurity interstitials
on the left and right boundaries of the solution domain. Combining
these expressions with solution (\ref{Homogeneous_solution}) of
the appropriate homogeneous diffusion equation, we obtain the
distribution of impurity interstitial concentration for the case
of the arbitrary boundary concentrations of impurity interstitial
atoms (see expression (\ref{Steady-state_solution2})).

\section{Simulation}

Using the system of Eqs. (\ref{Conservation law}) and
(\ref{Normalized equation}) and appropriate analytical solutions,
one can verify the correctness of  numerical calculations. This
system can be also used for modeling  hydrogen diffusion in
silicon. For the case of hydrogen diffusion, the quantities $C$
and $C^{AI}$ are the concentration of trapped (immobile) hydrogen
atoms and concentration of the neutral mobile hydrogen
interstitials, respectively. As example, in Fig. 1 the results of
simulation of hydrogen diffusion in silicon obtained on the basis
of analytical solution (\ref{Main integral}) and
(\ref{Steady-state_solution1D}) are presented. For comparison, the
experimental data of \cite{Sopori_96} are used. In
\cite{Sopori_96} deuterium was introduced in the laser
recrystallized silicon ribber by high-fluence low-energy (1.5 keV)
ion implantation at a temperature of 250 $^{\circ}$C. The
deuterium concentration profile presented in Fig. 1 was measured
by secondary ion mass spectrometry (SIMS). As can be seen from
Fig. 1, the  analytical solution obtained provides good agreement
with the experimental data \cite{Sopori_96}. The following values
of simulation parameters were used to fit the calculated curve to
the experimental deuterium profile: $Q$ = 4.32$\times$10$^{15}$
cm$^{-2}$; $p^{AI}$ = 0.253; $R_{P}$ = 0.035 $\mu$m; $\triangle
R_{P}$ = 0.028 $\mu$m; $l^{AI}$ =  0.31 $\mu$m. Here $Q$ is the
fluence of implanted hydrogen. Thus, it follows from the value of
the fitting parameter $p^{AI}$ that approximately 25\% of the
implanted deuterium atoms occupy interstitial positions. Migration
of these nonequilibrium interstitial atoms results in the
formation of extended ``tail" on the deuterium concentration
profile.

\begin{figure}[!ht]
\centering {
\begin{minipage}[ht]{12.0 cm}
{\includegraphics[scale=1.0]{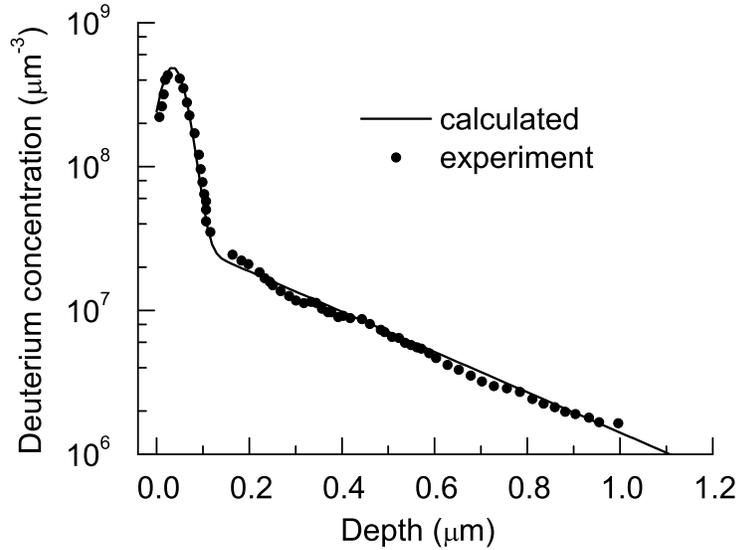}}
\end{minipage}
}

\caption{Deuterium concentration profile (solid line) calculated
for the case of ``hot'' ion implantation at a temperature of 250
$^{\circ}$C. The experimental data (dots) are taken from Sopori
{\it et al.} \cite{Sopori_96}. \label{Hydrogen}}
\end{figure}

In Fig. 2 the beryllium concentration profile after annealing
calculated on the basis of analytical solution (\ref{Main
integral}), (\ref{Integral_1}), (\ref{Integral_2}), and
(\ref{Integral_3}) is shown. For comparison, the experimental data
of \cite{Ihaddadene_04} are used. In Ref. \cite{Ihaddadene_04}
beryllium diffusion in InGaAs/InP during rapid annealing at a
temperature of 900 $^{\circ}$C for 30 s was investigated. In the
experiment under consideration, InP/InGaAs heterostructures were
grown by gas-source molecular beam epitaxy onto semi-insulating
$\langle100\rangle$ InP substrates. At first a 0.1 $\mu$m InP
buffer layer was grown, followed by 0.5 $\mu$m undoped InP and
then 0.2 $\mu$m Be-doped In$_{0.53}$Ga$_{0.47}$As layer with a
doping level of 3$\times$10$^{7}$ $\mu$m$^{-3}$ was grown.
Finally, an undoped InP layer of 0.5 $\mu$m was grown on the Be
doped layer. Then, a post-growth rapid thermal annealing was
performed in a halogen lamp furnace for 30 s. Beryllium profile
measurements were made with SIMS. The measurements confirmed that
the as-grown beryllium concentration profile is a step function,
similar to the function (\ref{Generation2}).

\begin{figure}[!ht]
\centering {
\begin{minipage}[ht]{12.0 cm}
{\includegraphics[scale=1.0]{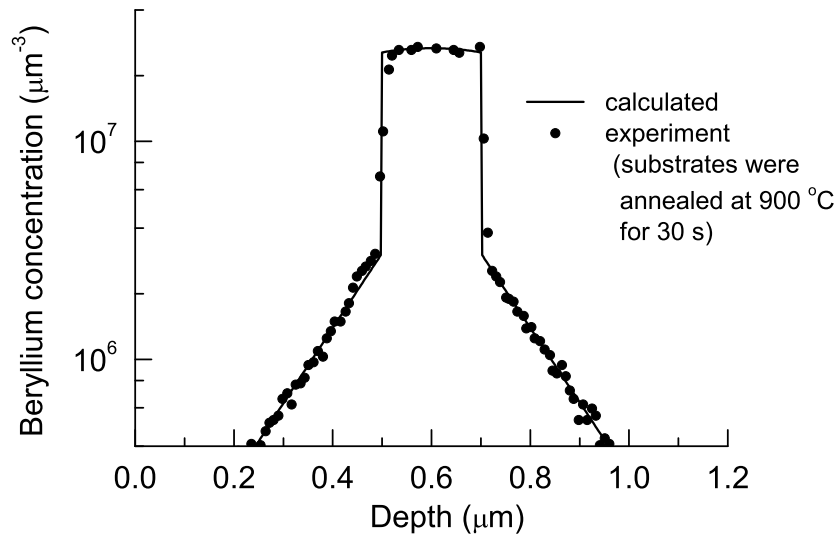}}
\end{minipage}
}

\caption{Beryllium concentration profile (solid line) calculated
for the case of rapid thermal annealing (30 s at a temperature of
900 $^{\circ}$C) of the InP/InGaAs heterostructures. The
experimental data (dots) are taken from Ihaddadene-Lenglet {\it et
al.} \cite{Ihaddadene_04}. \label{Beryllium}}
\end{figure}

To calculate the beryllium concentration profile after annealing
it is supposed that during thermal treatment a part of
substitutionally dissolved impurity atoms is transferred into
interstitial positions. If the generation rate of beryllium
interstitials is proportional to the impurity concentration, we
can use  function (\ref{Generation2}) to describe the distribution
of the interstitial generation rate. It is also supposed that the
impurity interstitial concentrations at the surface and in the
bulk of a semiconductor are equal to zero. Then, the beryllium
concentration profile after annealing is described by  analytical
expressions (\ref{Conservation law}), (\ref{Integral_1}),
(\ref{Integral_2}), and (\ref{Integral_3}).

As can be seen from Fig. 2, the calculated curve is in  good
agreement with the experimental data of \cite{Ihaddadene_04}. The
following values of simulation parameters were used to fit the
calculated curve to the experimental  concentration profile of
beryllium: $C_{max}$ = 2.672.$\times$10$^{7}$ $\mu$m$^{-3}$;
$p^{AI}$ = 0.254; $x_{L}$ = 0.5 $\mu$m; $x_{R}$ = 0.7 $\mu$m;
$l^{AI}$ = 0.127 $\mu$m. Here $C_{max}$ is the maximum
concentration of substitutionally dissolved beryllium after
annealing. It follows from the value of the fitting parameter
$p^{AI}$ that, as in the case of ``hot" ion implantation,
approximately 25\% of the beryllium atoms occupy the transient
interstitial positions. Migration of these nonequilibrium impurity
interstitial atoms results in the formation of two ``tails" on the
beryllium concentration profile after annealing.

\section{Conclusions}

The analytical solutions of the equations describing impurity
diffusion due to migration of nonequilibrium impurity
interstitials are obtained. Two representative cases were
investigated: i) interstitial impurity diffusion during ``hot" ion
implantation under reflecting boundary condition at the surface of
a semiconductor; ii) impurity interstitial diffusion from the
doped epitaxial layer under the conditions of constant
concentrations of interstitial impurity atoms at the surface and
in the bulk.

Using these solutions, we can verify the correctness of the
approximate numerical calculations obtained by the codes intended
for simulation of diffusion processes used in  fabrication of
semiconductor devices. Moreover, it is possible to carry out an
analytical simulation of certain diffusion processes. As an
example, hydrogen diffusion in silicon during high-fluence
low-energy deuterium implantation at a temperature of 250
$^{\circ}$C and beryllium diffusion from a doped epi-layer during
rapid thermal annealing of InP/InGaAs heterostructures at a
temperature of 900 $^{\circ}$C were investigated. The impurity
concentration profiles calculated on the basis of the analytical
solutions obtained agree well with the experimental data. Due to
comparison with experiment, the values of the parameters
describing interstitial impurity diffusion have been derived. It
was obtained that for the processes under consideration
approximately 25\% of the impurity atoms occupied  transient
interstitial positions. The average migration lengths of impurity
interstitials are 0.31 and 0.127 $\mu$m for deuterium in silicon
and beryllium in InP, respectively.

\end{document}